\documentclass[12pt]{article}

\usepackage{amsmath,amssymb,amsfonts,amsbsy}
\usepackage{graphicx}
\usepackage{wrapfig}

\begin{document}
\addcontentsline{toc}{subsection}{{Measurements of transverse spin effects 
in the forward region with the STAR detector}\\
{\it L.V. Nogach}}

\setcounter{section}{0}
\setcounter{subsection}{0}
\setcounter{equation}{0}
\setcounter{figure}{0}
\setcounter{footnote}{0}
\setcounter{table}{0}

\begin{center}
\textbf{ MEASUREMENTS OF TRANSVERSE SPIN EFFECTS \\ IN THE FORWARD REGION WITH THE STAR DETECTOR
}

\vspace{5mm}

L.V.~Nogach$^{\dag}$, for the STAR collaboration

\vspace{5mm}

\begin{small}
  \emph{Institute of High Energy Physics, Protvino, Russia} \\
  $\dag$ \emph{E-mail: Larisa.Nogach@ihep.ru}
\end{small}
\end{center}

\vspace{0.0mm} 

\begin{abstract}
Measurements by the STAR collaboration of the cross section and transverse single 
spin asymmetry (SSA) of neutral pion production at large Feynman x ($x_F$) 
in pp-collisions at $\sqrt{s}=200$ GeV were reported previously. The $x_F$ 
dependence of the asymmetry can be described by phenomenological models 
that include the Sivers effect, Collins effect or higher twist contributions 
in the initial and final states. Discriminating between the Sivers and 
Collins effects requires one to go beyond inclusive $\pi^0$ measurements. 
For the 2008 run, forward calorimetry at STAR was significantly 
extended. The large acceptance of the Forward Meson Spectrometer (FMS) 
allows us to look at heavier meson states and $\pi^0-\pi^0$ correlations.  
Recent results, the status of current analyses and near-term plans 
will be discussed.
\end{abstract}

\vspace{7.2mm} 

Contrary to simple perturbative QCD (pQCD) predictions, measurements 
of inclusive pion production in polarized proton-proton collisions at 
center-of-mass energies ($\sqrt{s}$) up to 20~GeV found large transverse 
single spin asymmetries ($A_N$) \cite{argonne,e704}. Measurements with
the Solenoidal Tracker at RHIC (STAR) detector confirmed that the $\pi^0$
asymmetry survives at $\sqrt{s}=$200~GeV and grows with increasing 
Feynman x ($x_F=2p_L/\sqrt{s}$) \cite{fpd1,fpd2}. Similar large spin effects
have recently been found in electron-positron and semi-inclusive deep
inelastic scattering \cite{belle,hermes}. Significant developments
in theory in the past few years suggest common origins for these effects, but 
the large $A_N$ in inclusive pion production is not yet fully understood.

A number of phenomenological models extend pQCD by introducing parton intrinsic 
transverse momentum ($k_T$) and considering correlations between parton $k_T$ 
and proton spin in the initial state (Sivers effect \cite{sivers}) or final 
state interactions of a transversely polarized quark fragmenting into a pion 
(Collins effect \cite{collins}). These models can explain the observed 
$x_F$ dependence of $A_N$, but their expectations of decreasing asymmetry 
with increasing pion transverse momentum ($p_T$) is not confirmed by the 
experimental data \cite{fpd2}. According to the current theoretical 
understanding both the Sivers and Collins mechanisms contribute to 
$\pi^0$ $A_N$ and discriminating between the two should be possible by 
measuring the asymmetry in direct photon production or forward jet 
fragmentation. 

Hadron production in the forward region in $pp$ collisions probes large-$x$ 
quark on low-$x$ gluon interactions and is a natural place to study spin 
effects. First measurements of that type at RHIC were done with the STAR 
Forward Pion Detector (FPD), a modular electromagnetic calorimeter placed 
at pseudorapidity $\eta\sim$~3-4 \cite{fpd1}. Prior to the 2008 run, forward 
calorimetry at STAR was extended with the FMS which replaced the 
FPD modules on one side (west) of the STAR interaction region. The FMS is 
a matrix of 1264 lead glass cells. It provides full azimuthal coverage in 
the region $2.5<\eta<4.0$ and has $\sim$20 times larger acceptance than the 
FPD. In addition to the study of inclusive $\pi^0$ production, this allows 
us to reconstruct heavier mesons and to look at ``jet-like'' events and 
$\pi^0-\pi^0$ correlations. 

The 2008 run at RHIC with transversely polarized proton beams measured an 
integrated luminosity of $\sim$7.8 $pb^{-1}$ at average beam polarization 
$P_{beam}\sim$45\%. The first step in the analysis of FMS data was to look 
at the inclusive $\pi^0$ asymmetry to make a point of contact with prior FPD 
measurements. $A_N$($x_F$) was found to be comparable to the previous 
results \cite{spin08}. The 2$\pi$ coverage in azimuthal angle ($\phi$) 
made possible a study of the $cos\phi$ dependence of the asymmetry, which 
is well described by a linear function, as expected \cite{spin08}. 
The FMS also allowed us to extend the measured $\pi^0$ $p_T$ region to 
$\sim$6 GeV/$c$ and added new data to investigate the $p_T$ dependence 
of $A_N$ \cite{cipanp09}.

\begin{figure}[b!]
  \centering
  \begin{tabular}{cc}
    \includegraphics[width=80mm]{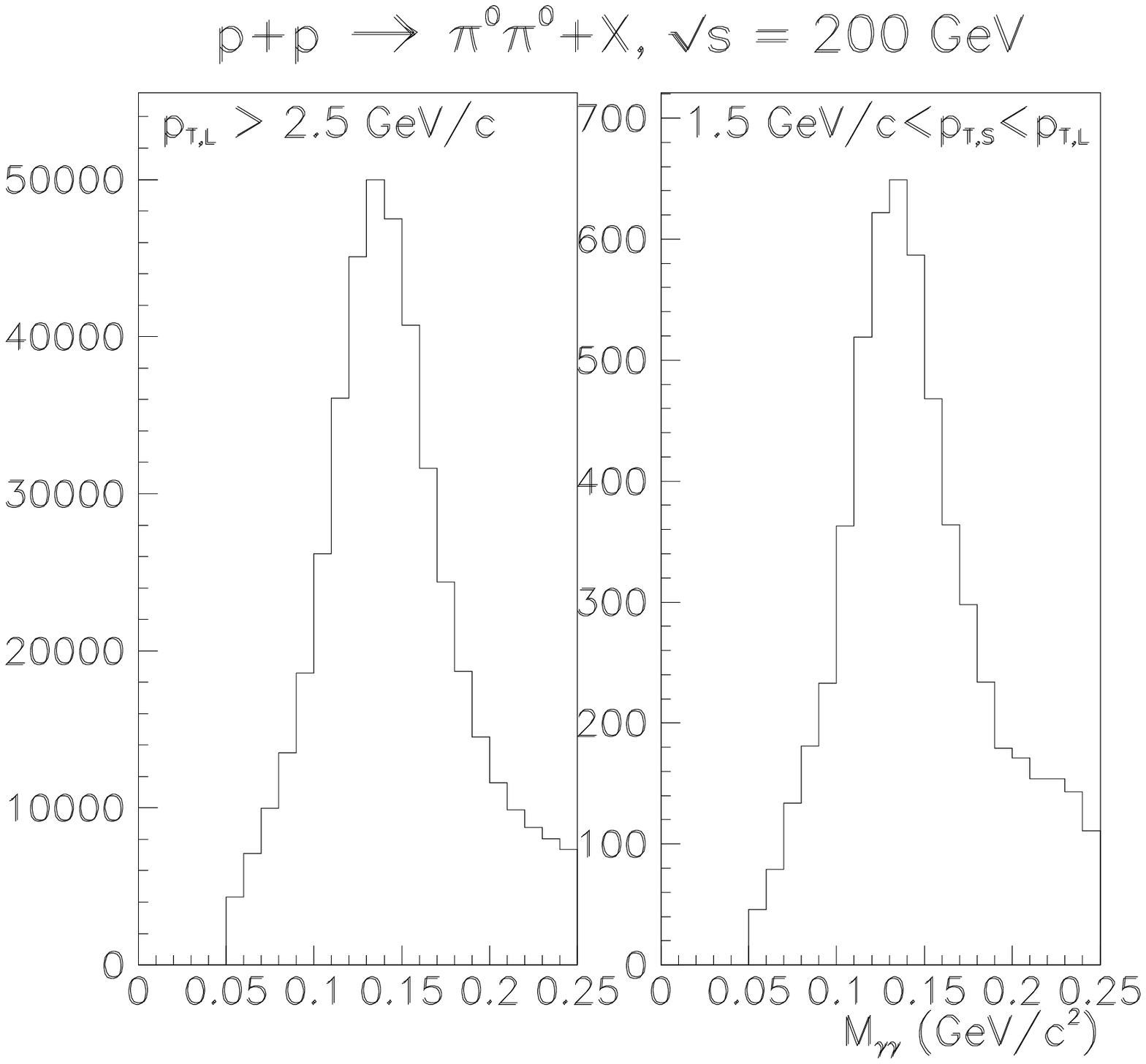} &
    \includegraphics[width=80mm]{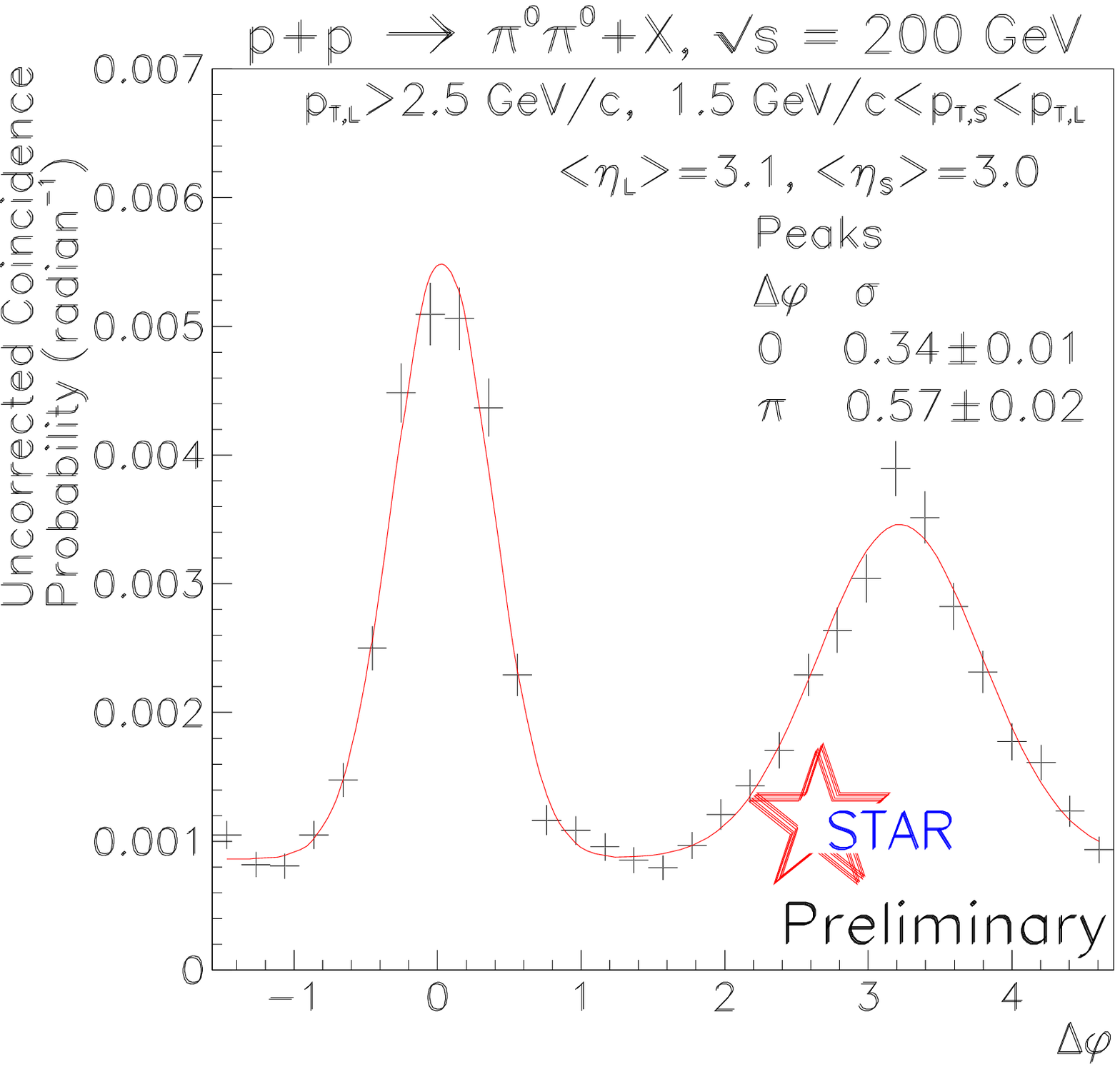} \\
    \textbf{(a)} & \textbf{(b)} 
  \end{tabular}
  \caption{\footnotesize
    \textbf{(a)} Invariant mass distributions for leading (left) and
                 subleading (right) neutral pions.  
    \textbf{(b)} Azimuthal correlations between two neutral pions in the FMS.
  }
  \label{fig_pipi}
\end{figure}

Another use of the large FMS acceptance was a reconstruction of forward neutral
pion pairs from $pp$ collisions. This analysis followed a few steps:
\begin{itemize}
\item[-] All photons in the FMS were reconstructed and a list was made of 
those satisfying the condition that the cluster energy was above 2 GeV and 
a fiducial volume requirement.
\item[-] $\pi^0$ reconstruction considered all possible two-photon combinations; 
a pion was identified if the di-photon invariant mass was between 0.05 and
0.25 GeV/$c^2$. The ``leading'' pion was required to have transverse momentum 
$p_{T,L} > 2.5$~GeV/$c$. The ``subleading'' pion was formed from the 
remaining photons and satisfied the requirement 1.5 GeV/$c < p_{T,S} < p_{T,L}$.
\item[-] The difference in azimuthal angle between the leading and subleading
pions ($\Delta\phi$) was calculated and the event distribution 
$dN/d\Delta\phi$, normalized by the number of leading pions, was plotted.   
\end{itemize}
The invariant mass for both pions is presented in Fig.~\ref{fig_pipi}a. 
Coincidence probability as a function of $\Delta\phi$ is shown in 
Fig.~\ref{fig_pipi}b. The distribution has two clear peaks corresponding to 
``near-side'' and ``away-side'' correlations, and is fitted by two Gaussian 
functions plus a constant background. The correlations are not yet normalized.
Simulations to obtain efficiency corrections are under way.  

The above analysis can be extended to look at the correlations sorted by
the spin state of the colliding protons. Back-to-back ($\Delta\phi \approx \pi$) 
di-hadron measurements can provide access to the Sivers function, 
as suggested in \cite{boer}, assuming that the neutral pions serve as jet 
surrogates. Near-side hadron correlations can provide sensitivity
to the Collins fragmentation function and transversity. 

Near-term plans for forward physics at STAR include measurements of the cross 
section and transverse SSA in inclusive $\pi^0$ production in polarized 
proton-proton collisions at $\sqrt{s}=500$~GeV and a proposal to add 
a Forward Hadron Calorimeter (FHC) behind the FMS. 

The 2009 run at RHIC was the first physics run at $\sqrt{s}=500$~GeV. STAR 
sampled 10~$pb^{-1}$ of data with longitudinally polarized beams at this energy. 
A first look at the FPD data shows that with the lead-glass matrices alone, 
$\pi^0$ events can be reconstructed up to $x_F\sim~0.25$. Each FPD module 
also includes a two-plane scintillation shower maximum detector that provides 
essential data to separate photons from decays of pions at higher $x_F$. 
Measurements with transversely polarized beams at $\sqrt{s}=500$~GeV are 
tentatively planned for the 2011 run. 

\begin{wrapfigure}[16]{R}{80mm}
  \centering 
  \vspace*{-4mm} 
  \includegraphics[width=75mm]{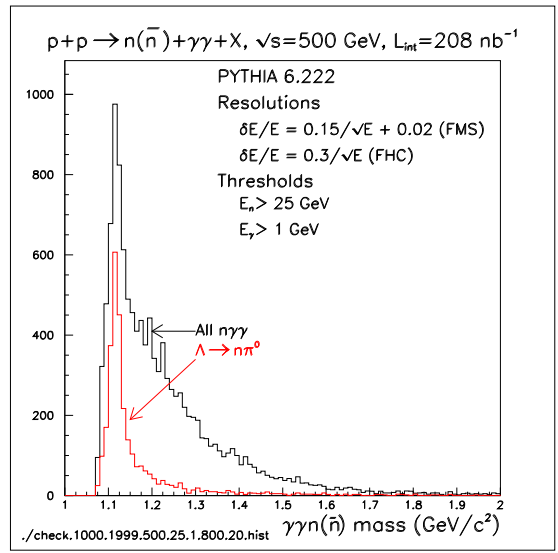}
  \vspace*{-18mm} 
  \caption{\footnotesize
Three-cluster mass distributions from the simulation of 
$pp\rightarrow n(\bar n)+2\gamma+X$ events.}
  \label{fig_lambda}
\end{wrapfigure}

The proposed addition of the FHC (two matrices of 9$\times$12 lead-scintillator  
detectors) to the STAR detector is motivated by the following physics goals:
\begin{itemize}
\item[-] to measure the transverse SSA for full jets that should allow 
to isolate the contribution from the Sivers mechanism to the observed 
$\pi^0$ asymmetry;
\item[-] measurements of polarization transfer coefficients through $\Lambda$
polarization in the $\Lambda\rightarrow n\pi^0$ channel to test pQCD 
predictions.
\end{itemize}
PYTHIA simulation of $pp\rightarrow n(\bar n)+2\gamma+X$ events in 
the FMS+FHC have been done using a fast event generator method. Reconstructed
mass for all $n\gamma\gamma$ events and for the $\Lambda\rightarrow n\pi^0$
process is shown in Fig.~\ref{fig_lambda}.

In summary, precision measurements of $\pi^0$ $A_N$ with the FPD allow
for a quantitative comparison with theoretical models. The FMS allows
us to go beyond inclusive $\pi^0$ production to heavier mesons, ``jet-like'' 
events and particle correlation studies. Measurements of $A_N$ for direct
photons or jets are needed to disentangle the dynamical origins. Prospects 
for the more distant future include the development of a RHIC experiment 
to measure transverse SSA for Drell-Yan production of dilepton pairs.

\end{document}